\documentclass{PoS}

\def\beqn{\begin{eqnarray}} \def\eeqn{\end{eqnarray}}
\def\beq{\begin{equation}} \def\eeq{\end{equation}}
\def\nn{\nonumber}
\def\td#1{\tilde{\delta}\left(#1\right)}

\def\qb{\mathbf{q}}

\def\ii{\imath 0}

\title{Analysis of singularities and the four-dimensional representation of physical observables within the LTD formalism}

\ShortTitle{Analysis of singularities and 4D representations within the LTD formalism}

\author{\speaker{German F. R. Sborlini}$^{\ a}$\\
        $^a$Instituto de F\'{\i}sica Corpuscular, Universitat de Val\`{e}ncia -- 
Consejo Superior de Investigaciones Cient\'{\i}ficas, Parc Cient\'{\i}fic, E-46980 Paterna, Valencia, Spain.\\
        E-mail: \email{german.sborlini@unimi.it}}

\abstract{In the past years, we have been developing a novel technique, called Four-Dimensional Unsubtraction (FDU) which aims to obtain purely four-dimensional representations of the matrix elements contributing to physical observables. In this talk, we describe the application of the loop-tree duality (LTD) theorem to represent loop amplitudes in terms of tree-level like objects, focusing on the origin of possible singularities of scattering amplitudes. In particular, we analyze the regions responsible of infrared and threshold singularities. With this information, we aim to extend the FDU formalism to NNLO and beyond.}

\FullConference{
European Physical Society Conference on High Energy Physics - EPS-HEP2019 -\\
			10-17 July, 2019\\
			Ghent, Belgium}

\begin{document}

\section{Introduction}
\label{sec:introduction}
When dealing with scattering amplitudes in the context of perturbative Quantum Field Theories (pQFT), it is usual to find ill-defined mathematical expressions. In particular, tree-level amplitudes are rational functions of the external momenta whose poles indicate the presence of infrared (IR) or threshold singularities. On the other hand, loop amplitudes also posses ultraviolet (UV) divergences and thresholds that originate branch-cuts. The important observation is that scattering amplitudes can be reconstructed from their singular structure, which indicates that most of the physical information of these objects is contained within the problematic regions.

In this talk, we propose to study the structure of loop amplitudes through the application of the Loop-Tree Duality (LTD) theorem \cite{Catani:2008xa,Rodrigo:2008fp}. This method has several advantages compared to other approaches because the integration domain is transformed into an Euclidean space, thus reducing the treatment of loop amplitudes to tree-level-like objects defined over an extended phase-space. In this way, the analysis of singularities is more transparent and allows a direct physical interpretation in terms of real particles.

The outline of this article is the following. In Sec. \ref{sec:LTDintro}, we very briefly review the main notation required for implementing the LTD approach, as well as some basic formulae. Then, we directly jump to the analysis of IR singularities for loop amplitudes, through an explicit example. This is discussed in Sec. \ref{sec:IRsingularities}, where we focus on the location of the problematic regions within the integration domain. After that, we enter into the analysis of threshold singularities in Sec. \ref{sec:Thresholds}. We summarize the conclusions of a very recent paper \cite{Aguilera-Verdugo:2019kbz}, emphasizing the advantages of the LTD approach to deal with generic multi-loop multi-leg processes. Finally, we present the conclusions and future research directions in Sec. \ref{sec:conclusions}.

\section{Basics of the LTD theorem}
\label{sec:LTDintro}
It is a well-known fact that virtual amplitudes can be decomposed by performing cuts on internal lines. In the early sixties, the Feynman Tree theorem (FTT) established that any loop integral can be expressed as the sum of all possible $l$-cuts, with $l$ running over all the available internal lines. This idea lead to several developments in the following years, such as the unitarity based methods. Within the FTT, all the cut contributions preserve the original prescriptions. However, the number of possible cuts rapidly increases with the number of loops and external legs. An alternative approach consists in applying the Cauchy's residue theorem on the energy component of the loop momenta, thus imposing only one on-shell condition for each independent loop momenta: this is the Loop-Tree Duality (LTD) theorem. From a practical point of view, it can be stated that the number of simultaneous cuts is equal to the number of loops. Apart from that, the prescription of the dual components  within LTD has to be modified in order to capture the information contained inside the multiple-cut contributions given by the FTT. 

Explicitly, at one-loop, a generic $N$-particle scalar integral is written as
\beqn
L^{(1)}(p_1,\ldots,p_N) &=& \int_\ell \, \prod_{i=1}^N G_F(q_i) \equiv \sum_{i=1}^N \int_\ell \tilde{\delta}(q_i) \prod_{j=1, j\neq i}^N \, G_D(q_i;q_j) \, , 
\label{eq:MasterLTD1}
\eeqn
where the so-called \emph{dual} propagator is defined according to
\beq
G_D(q_i;q_j) = \frac{1}{q_j^2-m_j^2-\imath 0 \, \eta(q_j-q_i)} \, ,
\label{eq:GDdefinition}
\eeq  
being $\eta$ an arbitrary space-like vector. It is worth appreciating that $\tilde{\delta}(q_i)=2\pi\, \imath \, \delta(q_i^2-m_i^2)\theta(q_{i,0})$, which implies that the dual contributions are integrated over on-shell forward hyperboloids.

The formula can be generalized to the multi-loop case, and also allows to deal with Feynman integrals involving higher-powers \cite{Bierenbaum:2010cy} of the propagators (for instance, those associated to self-energy insertions beyond the two-loop level). For instance, at two-loop, we can write \cite{Bierenbaum:2012th,Driencourt-Mangin:2019aix}
\beqn
\nn && {\cal L}^{(2)}(p_1,\ldots,p_N) = \int_{\ell_1}\int_{\ell_2} \prod_{i \in \alpha} \, G_F(q_i) 
\nn \\ &\equiv& \int_{\ell_1}\int_{\ell_2} \left[ G_D(\alpha_1) \, G_D(\alpha_2\cup \alpha_3) 
+ G_D(-\alpha_2\cup \alpha_1) G_D(\alpha_3) - G_D(\alpha_1) \,G_F(\alpha_2) \, G_D(\alpha_3) \right] ~,
\label{eq:LTDtwoloop}
\eeqn
where the internal momenta $q_i = \ell_1+k_i$, $q_j = \ell_2 + k_j$ and $q_k = \ell_1+ \ell_2 + k_k$,
are classified into three different sets, $i \in \alpha_1$, $j \in \alpha_2$ and $k \in \alpha_3$, with $\alpha=\alpha_1\cup\alpha_2\cup\alpha_3$. More details about the momenta assignation is available in Ref. \cite{Driencourt-Mangin:2019aix}.


\section{Location of IR singularities and FDU approach}
\label{sec:IRsingularities}
Let us consider the dual representation of a generic one-loop scalar integral given in Eq. (\ref{eq:MasterLTD1}). If we center in the $i$-th dual contribution, we can appreciate that it involves the product of dual propagators $G_D(q_i;q_j)$ with $j \neq i$, integrated over the forward hyperboloid associated to the positive energy solution of $G_F(q_i)^{-1}=0$. Inside this region, it is possible that one or more internal momenta become on-shell, thus leading to the appearance of singularities at integrand level. These singularities can be due to intersection of hyperbolodois or a complete overlap of them. The first situation originates threshold singularities, whilst the second one can only take place when dealing with massless states and is responsible of the infrared (IR) divergent structure of the integral \cite{Buchta:2014dfa,Hernandez-Pinto:2015ysa}. We will return to thresholds in the next section.

\begin{figure}[htb]
\begin{center}
\includegraphics[width=0.95\textwidth]{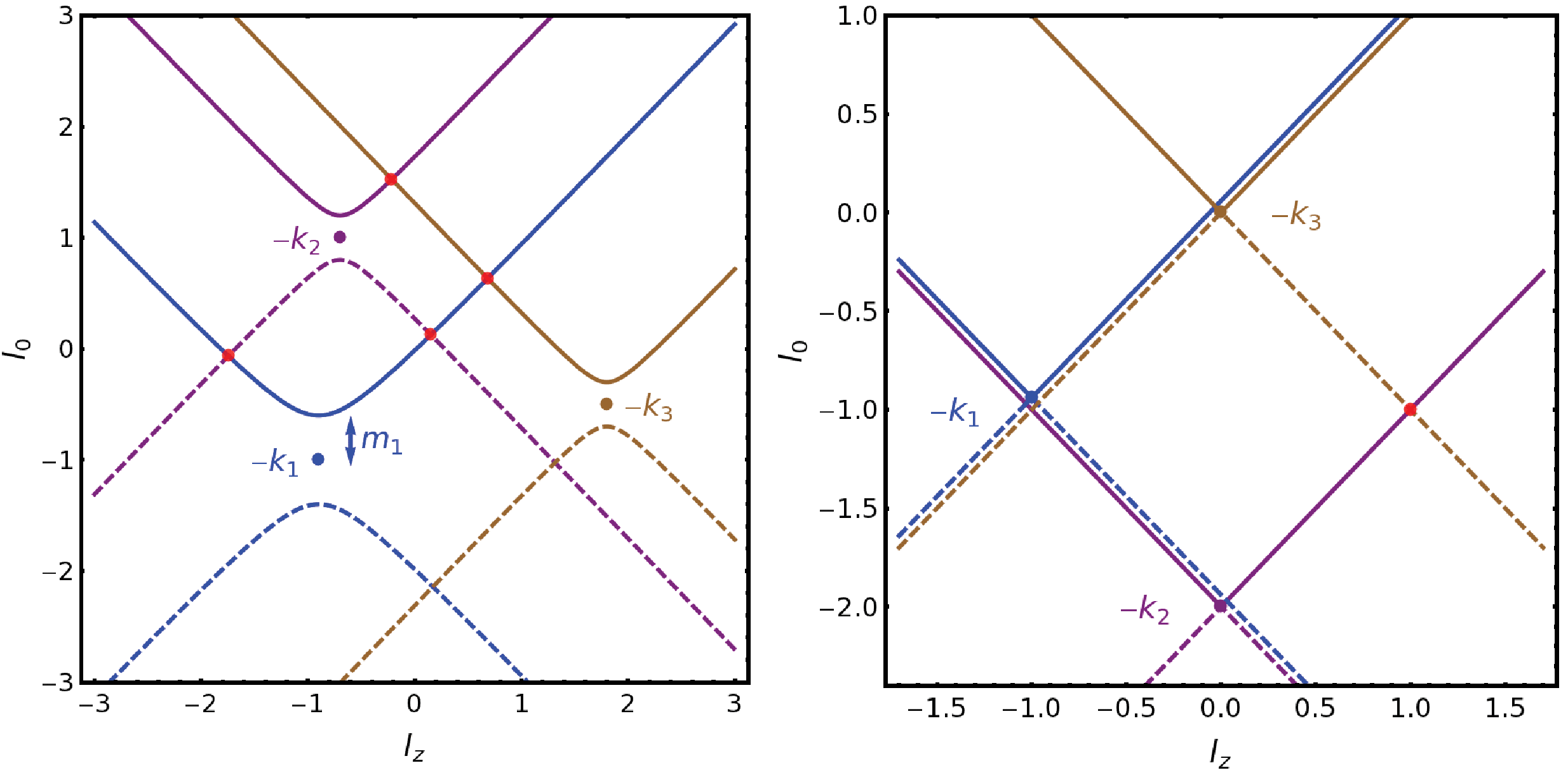}
\caption{Graphical representation of the on-shell hyperboloids associated to the integration region for the different dual contributions. Their intersections determine the location of IR and threshold singularities within the integration domain. In the left panel, internal masses prevent the overlap of hyperboloids and there are not IR singularities, but threshold ones. In the right panel, hyperboloids degenerate to light-cones: forward-backward intersections originate the IR singular structure.
\label{fig:IRsingularities}}
\end{center}
\end{figure}

In Fig. \ref{fig:IRsingularities} we show the previously mentioned situations for the particular case of a three-point scalar function. In the left panel, internal states are massive, thus leading to non-overlapping hyperboloids. In this case, they could partially intersects, and originate quasi-collinear singularities (that manifest through logarithmically enhanced contributions). In the right panel, we consider the massless triangle: we can appreciate that the forward light-cone of $I_1$ overlaps with the backward light-cone of $I_3$ and the forward light-cone of $I_2$ overlaps with the backward one of $I_1$. These intersections lead to IR soft/collinear singularities, which are contained within a compact region of the integration domain. 

The last assertion is one of the main ingredients of the so-called Four-Dimensional Unsubtraction (FDU) framework. According to several well-known theorems, such as the Kinoshita-Lee-Nauenberg theorem for QCD, the sum over all degenerated configurations allows to define finite IR-safe observables. From the practical point of view, this implies summing over virtual and real contributions. The LTD theorem transforms the virtual into dual terms, but still it is necessary to perform a proper integrand level combination with the real-emission part. In order to do that, we need to define proper kinematical mappings to generate the real phase-space. The fact that IR singularities are confined into a compact region for both the dual and the mapped-real contributions guarantees the point-by-point cancellation of IR singularities: once the integrand-level expressions are locally regularized, we can safely work in four spcae-time dimensions.

More details about this method are available in Refs. \cite{Hernandez-Pinto:2015ysa,Sborlini:2015uia,Sborlini:2016fcj,Sborlini:2016gbr,Sborlini:2016hat,Sborlini:2016saj,Sborlini:2016bod,Sborlini:2017nee,Rodrigo:2018jme}.

\section{Thresholds and unphysical singularities}
\label{sec:Thresholds}
The LTD approach offers a very suitable framework to understand the origin of the singular structure of virtual contributions, performing the whole analysis in four space-time dimensions. This is mainly due to the fact that Minkowski integration domain is transformed into an Euclidean one, which resembles a physical phase-space integral. So, let us apply this to analyze the singular behaviour of one-loop amplitudes; a complete study at two-loops and beyond is presented in Ref. \cite{Aguilera-Verdugo:2019kbz}.

The starting point consists in the definition of the integrand functions
\beqn
{\cal S}^{(1)}_{ij} &=& (2\pi \imath)^{-1}\ G_D(q_i;q_j) \,  \td{q_i} + (i \leftrightarrow j)~,
\label{eq:OneLoopS1ij}
\\ {\cal S}^{(1)}_{ijk} &=& (2\pi \imath)^{-1}\ G_D(q_i;q_j)G_D(q_i;q_k) \,  \td{q_i} + {\rm perms.}~,
\label{eq:OneLoopS1ijk} 
\eeqn
which are associated to the sum of all the possible single-cuts for bubble and triangle integrals, respectively. Given $q_{l,0}^{(+)} = \sqrt{\qb_l^2+m_l^2}$, with $l\in\{i,j\}$, we can introduce the set of variables
\beqn
\lambda^{\pm\pm}_{ij} &=& \pm q_{i,0}^{(+)}  \pm q_{j,0}^{(+)}  + k_{ji,0} \, , 
\label{eq:LambdaVariables}
\eeqn
with $k_{ji}$ a combination of momenta\cite{Aguilera-Verdugo:2019kbz}. Since the denominators of different dual propagators consist in products of different $\lambda^{\pm\pm}_{ij}$, the limits $\lambda^{\pm\pm}_{ij} \to 0$ indicate the location of the singular points of the integrands. In consequence, without losing generality, we can focus on the solutions of the equations
\beq
\lambda^{\pm\pm}_{ij} = 0 \ ,
\label{eq:EquationMaster}
\eeq
and explore the conditions allowing such kinematical configurations.

For the moment, let us consider the function ${\cal S}^{(1)}_{ij}$. Its singular structure depends on the distance of the external momenta measured in the four-dimensional space-time. In this case, there are only two independent limits:
\beq
\lambda^{++} \to 0 \ \ \ \ {\rm and} \ \ \ \ \lambda^{+-} \to 0 \, .
\eeq
The solution $\lambda^{++}_{ij} = 0$ requires
\beq
k_{ji}^2 -(m_i+m_j)^2 \geq 0 \, ,
\eeq
which implies a time-like separation between the momenta, i.e. they are causally connected. This region is responsible of physical threshold singularities that fix the imaginary part of the corresponding Feynman integrals. By expanding ${\cal S}_{ij}^{(1)}$ around $\lambda_{ij}^{++}=0$, we find
\beqn
{\cal S}^{(1)}_{ij} &=& - \frac{\theta(-k_{ji,0}) 
 \theta( k_{ji}^2 - (m_i+m_j)^2)} 
{4 \, q_{i,0}^{(+)} q_{i,0}^{(+)} \, (\lambda^{++}_{ij}+\ii k_{ji,0})}
+{\cal O}\left((\lambda_{ij}^{++})^0\right)~,
\eeqn
where we can appreciate that the prescription is always $-\ii$ because $k_{ji,0}<0$. This is equivalent to fix the standard Feynman prescription, thus implying that LTD and FTT give the same contribution. Another interpretation of the limit $\lambda_{ij}^{++}\to0$ is possible by using hyperboloids. In fact, this situation corresponds to the intersection of a forward with a backward hyperboloid.

The other independent limit, namely $\lambda^{+-}_{ij} \to 0$, requires
\beq
k_{ji}^2 -(m_i-m_j)^2 \leq 0 \, ,
\eeq
i.e. a non-causal or space-like separation of the momenta. In this case, it is possible to demonstrate that an unphysical threshold is originated. However, the property
\beq
\lim_{\lambda_{ij}^{+-} \to 0} q_{j,0}^{(+)} G_D(q_i;q_j) = - \lim_{\lambda_{ij}^{+-} \to 0} q_{i,0}^{(+)} G_D(q_j;q_i) \,
\eeq
is valid because of the change of sign within the dual prescription. This implies that the potential singular behavior is healed when adding the different dual contributions, thus canceling in the final result. In other terms, ${\cal S}^{(1)}_{ij}={\cal O}((\lambda_{ij}^{+-})^0)$. At this point, we can appreciate that the result is completely consistent with the FTT decomposition, where the cancellation of unphysical singularities occurs after adding all the possible cuts (even multiple cuts). The advantage of the LTD approach is that such cancellation only requires to consider single cuts with the proper modified prescription.

As we did with physical thresholds, the LTD approach has the advantage of allowing a geometrical interpretation of the cancellation of unphysical or spurious singularities. They are originated in the intersection of two forward (backward) hyperboloids. In this way, the same potential singularity manifest in two separated dual terms, but they contribute with an opposite sign.


\subsection{Anomalous thresholds}
\label{ssec:anomalous}
Finally, we would like to highlight the advantages of the LTD formalism to provide an alternative description of anomalous thresholds. These singular configurations were first studied in the sixties, but a mathematically rigorous treatment is not offer within the most diffused literature. At one-loop level, anomalous thresholds manifest when more than two propagators become simultaneously on-shell. Using the function ${\cal S}^{(1)}_{ijk}$ introduced in Eq. (\ref{eq:OneLoopS1ijk}), we can consider the double limit
\beq
\lambda^{++}_{ij} \to 0 \ \ \ \ {\rm and} \ \ \ \  \lambda^{++}_{ik} \to 0 \, ,
\eeq
with $k_{ji,0} \leq 0$ and $k_{ki,0} \leq 0$. The geometrical interpretation corresponds to the triple intersection of forward (backward) hyperboloids: cross-cancellations will take place among the different dual contributions. Here, it is crucial to take into account the change of sign of the dual prescription to guarantee the cancellation.

However, a physical effect could be generated due to the intersection of two forward (backward) and one backward (forward) hyperboloids. For more details about anomalous thresholds within the LTD approach at one-loop and beyond, we refer the reader to Ref. \cite{Aguilera-Verdugo:2019kbz}.

\section{Outlook and conclusions}
\label{sec:conclusions}
In this article, we summarize some recent developments on the analysis of the singular structure of scattering amplitudes within the LTD approach. We focus on the advantages offered by our technique, since all the analysis is performed at the level of dual integrals defined in Euclidean spaces. This allowed us to locate the regions responsible of the physical IR singularities, and achieve a fully local cancellation when adding the mapped-real corrections \cite{Sborlini:2016gbr,Sborlini:2016hat}. In consequence, a natural four-dimensional representation of any IR-safe physical observables emerges from our formalism.

Also, we have recently exploted our method to describe the origin of causal and anomalous thresholds, at one-loop level and beyond \cite{Aguilera-Verdugo:2019kbz}. We demonstrated that the dual prescription is crucial to achieve a proper cancellation of unphysical and spurious singularities. Moreover, we paved the way for a deeper understanding of the causal structure of scattering amplitudes and their singularities at higher-loop orders, by transforming Feynman integrals into phase-space ones through the LTD theorem.

\section*{Acknowledgments}
\label{sec:Acknowledgements}
This work is supported by the Spanish Government (Agencia Estatal de Investigacion) and ERDF funds from European Commission (Grants No. FPA2017-84445-P and SEV-2014-0398), by Generalitat Valenciana (Grant No. PROMETEO/2017/053), by Consejo Superior de Investigaciones Cient\'ificas (Grant No. PIE-201750E021). The author also acknowledges the support by the COST Action CA16201 PARTICLEFACE.


\begin{thebibliography}{99}

\bibitem{Catani:2008xa}
  S.~Catani, T.~Gleisberg, F.~Krauss, G.~Rodrigo and J.~C.~Winter,
  JHEP {\bf 0809} (2008) 065
  [arXiv:0804.3170 [hep-ph]].

\bibitem{Rodrigo:2008fp}
  G.~Rodrigo, S.~Catani, T.~Gleisberg, F.~Krauss and J.~C.~Winter,
  Nucl.\ Phys.\ Proc.\ Suppl.\  {\bf 183} (2008) 262
  [arXiv:0807.0531 [hep-th]].

\bibitem{Aguilera-Verdugo:2019kbz}
  J.~J.~Aguilera-Verdugo, F.~Driencourt-Mangin, J.~Plenter, S.~Ram\'irez-Uribe, G.~Rodrigo, G.~F.~R.~Sborlini, W.~J.~Torres Bobadilla and S.~Tracz,
  arXiv:1904.08389 [hep-ph].
 
\bibitem{Bierenbaum:2010cy}
  I.~Bierenbaum, S.~Catani, P.~Draggiotis and G.~Rodrigo,
  JHEP {\bf 1010} (2010) 073
  [arXiv:1007.0194 [hep-ph]].

\bibitem{Bierenbaum:2012th}
  I.~Bierenbaum, S.~Buchta, P.~Draggiotis, I.~Malamos and G.~Rodrigo,
  JHEP {\bf 1303} (2013) 025
  [arXiv:1211.5048 [hep-ph]].

\bibitem{Driencourt-Mangin:2019aix}
  F.~Driencourt-Mangin, G.~Rodrigo, G.~F.~R.~Sborlini and W.~J.~Torres Bobadilla,
  JHEP {\bf 1902} (2019) 143
  [arXiv:1901.09853 [hep-ph]].
  
\bibitem{Buchta:2014dfa}
  S.~Buchta, G.~Chachamis, P.~Draggiotis, I.~Malamos and G.~Rodrigo,
  JHEP {\bf 1411} (2014) 014
  [arXiv:1405.7850 [hep-ph]].

\bibitem{Hernandez-Pinto:2015ysa}
  R.~J.~Hern\'andez-Pinto, G.~F.~R.~Sborlini and G.~Rodrigo,
  JHEP {\bf 1602} (2016) 044
  [arXiv:1506.04617 [hep-ph]].

\bibitem{Sborlini:2015uia}
  G.~F.~R.~Sborlini, R.~Hern\'andez-Pinto and G.~Rodrigo,
  PoS {\bf EPS-HEP2015} (2015) 479
  [arXiv:1510.01079 [hep-ph]].

\bibitem{Sborlini:2016fcj}
  G.~F.~R.~Sborlini,
  PoS {\bf RADCOR2015} (2016) 082
  [arXiv:1601.04634 [hep-ph]].

\bibitem{Sborlini:2016gbr}
  G.~F.~R.~Sborlini, F.~Driencourt-Mangin, R.~Hern\'andez-Pinto and G.~Rodrigo,
  JHEP {\bf 1608} (2016) 160
  [arXiv:1604.06699 [hep-ph]].

\bibitem{Sborlini:2016hat}
  G.~F.~R.~Sborlini, F.~Driencourt-Mangin and G.~Rodrigo,
  JHEP {\bf 1610} (2016) 162
  [arXiv:1608.01584 [hep-ph]].

\bibitem{Sborlini:2016saj}
  G.~F.~R.~Sborlini,
  PoS {\bf ICHEP2016} (2016) 658
  [arXiv:1611.05094 [hep-ph]].

\bibitem{Sborlini:2016bod}
  G.~F.~R.~Sborlini, F.~Driencourt-Mangin, R.~Hern\'andez-Pinto and G.~Rodrigo,
  PoS {\bf ICHEP2016} (2016) 353
  [arXiv:1611.04824 [hep-ph]].

\bibitem{Sborlini:2017nee}
  G.~F.~R.~Sborlini, F.~Driencourt-Mangin, R.~Hern\'andez-Pinto and G.~Rodrigo,
  PoS {\bf EPS-HEP2017} (2017) 547
  [arXiv:1710.04516 [hep-ph]].

\bibitem{Rodrigo:2018jme}
  G.~Rodrigo, F.~Driencourt-Mangin, G.~F.~R.~Sborlini and R.~J.~Hern\'andez-Pinto,
  PoS {\bf RADCOR2017} (2018) 013
  [arXiv:1801.04465 [hep-ph]].



\end{thebibliography}
\end{document}